\documentclass[aps,amsmath,amssymb,twocolumn,longbibliography, superscriptaddress]{revtex4-1}
\usepackage{graphicx}%
\graphicspath{{Figures/}}
\usepackage{dcolumn}%
\usepackage{bm}%
\usepackage{mathrsfs}
\usepackage{subfigure}
\usepackage{natbib}
\usepackage{amssymb}
\usepackage{multirow}
\usepackage{float}
\usepackage{color}
\usepackage{appendix}
\usepackage[dvipsnames]{xcolor}
\usepackage[colorlinks, allcolors=Blue]{hyperref}
\usepackage{amsmath}
\usepackage{booktabs}
\usepackage{xr}
\newcommand{\p}{\ensuremath \partial}

\newcommand{\Pe}{\ensuremath \mathrm{P\hspace{-1pt}e}}
\newcommand{\cIn}{c_\mathrm{in}}
\newcommand{\diff}{\ensuremath \mathrm{d}}
\newcommand{\Eqref}[1]{Eq.~\eqref{#1}}
\newcommand{\Eqsref}[1]{Eqs.~\eqref{#1}}
\newcommand{\Figref}[1]{Fig.~\ref{#1}}

\renewcommand{\tableofcontents}{}

\begin{document}

\title{Advection selects pattern in multistable emulsions of active droplets}

\author{Stefan Köstler}
\affiliation{Max Planck Institute for Dynamics and Self-Organization, Am Faßberg 17, 37077 Göttingen, Germany}
\affiliation{University of Göttingen, Institute for Theoretical Physics, Friedrich-Hund-Platz 1, 37077 Göttingen, Germany}

\author{Yicheng Qiang}
\affiliation{Max Planck Institute for Dynamics and Self-Organization, Am Faßberg 17, 37077 Göttingen, Germany}

\author{Guido Kusters}
\affiliation{Max Planck Institute for Dynamics and Self-Organization, Am Faßberg 17, 37077 Göttingen, Germany}

\author{David Zwicker}
\affiliation{Max Planck Institute for Dynamics and Self-Organization, Am Faßberg 17, 37077 Göttingen, Germany}

\begin{abstract}
Controlling the size of droplets, for example in biological cells, is challenging because large droplets typically outcompete smaller droplets due to surface tension. This coarsening is generally accelerated by hydrodynamic effects, but active chemical reactions can suppress it. We show that the interplay of these processes leads to three different dynamical regimes: (1) Advection dominates the coalescence of small droplets, (2) diffusion leads to Ostwald ripening for intermediate sizes, and (3) reactions finally suppress coarsening. Interestingly, a range of final droplet sizes is stable, of which one is selected depending on initial conditions. Our analysis demonstrates that hydrodynamic effects control initial droplet sizes, but they do not affect the later dynamics, in contrast to passive emulsions.
\end{abstract}

\maketitle

\tableofcontents

\section{Introduction}
\label{int}

Emulsions of multiple droplets provide spatial structures in biological cells~\cite{brangwynneGermlineGranulesAre2009,hymanLiquidLiquidPhaseSeparation2014,bananiBiomolecularCondensatesOrganizers2017,zwickerPhysicsDropletRegulation2025} and synthetic applications~\cite{jambon-puilletPhaseseparatedDropletsSwim2024,songSyntheticBiomolecularCondensates2025}.
In these examples, effective patterning requires controlled droplet sizes and positions.
These aspects are affected by physical processes such as diffusion, advection, and chemical reactions~\cite{zwickerPhysicsDropletRegulation2025}.
In particular, diffusion generally causes droplet coarsening, either because molecules diffuse from smaller to larger droplets (Ostwald ripening~\cite{ostwaldStudienUeberBildung1897,voorheesTheoryOstwaldRipening1985}) or because droplet diffusion leads to coalescence~\cite{tanakaNewMechanismsDroplet1997}.
Hydrodynamic advection generally accelerates coarsening~\cite{nikolayevNewHydrodynamicMechanism1996,tanakaNewMechanismsDroplet1997,huoHydrodynamicEffectsPhase2003}, whereas driven chemical reactions can suppress coarsening~\cite{wurtzChemicalReactionControlledPhaseSeparated2018,bressloffActiveSuppressionOstwald2020,sastreSizeControlOscillations2024,bauermannCriticalTransitionIntensive2025,bauermannTheoryReversedRipening2025}.
However, how the interplay of these processes organizes emulsions, and in particular selects length scales, is poorly understood.

\section{Results}
\subsection{Chemically active droplets couple diffusion, advection, and reactions}
\label{theo}

To unveil the interplay between diffusion, advection, and reactions, we study a minimal model of an emulsion described by an isothermal fluid comprising two chemical species A and B. 
We assume incompressibility, so that the system's state is described by only the number concentration $c(\boldsymbol{r},t)$ of species A, which evolves as %
\begin{align}
    \p_t c + \boldsymbol{v}\cdot\nabla c = \Lambda \nabla^2 \mu - k(c - c_0)
    \;,
    \label{eq:AdvReacCH}
\end{align}
where the left-hand side describes the material derivative with velocity field $\boldsymbol{v}$, whereas the right-hand side accounts for diffusion with constant mobility $\Lambda$ and a linear reaction with rate $k$ describing the conversion between A and~B.
This conversion reaction describes a driven reaction since it is not derived based on thermodynamic principles~\cite{Lefever1995,Kirschbaum2021,zwickerPhysicsDropletRegulation2025}.
In particular, the activity-dependent reaction equilibrium at $c=c_0$ is distinct from the thermodynamic equilibrium described by the free energy $F$ of the system.
In contrast, the diffusive flux is driven by gradients in the exchange chemical potential $\mu = \delta F/\delta c$, implying that it biases the system toward the free energy minimum.
To analyze the interplay of passive diffusion and active reactions, we consider a free energy of the Ginzburg--Landau form~\cite{weberPhysicsActiveEmulsions2019}
\begin{equation}
    F = \int \mathrm{d}V \left[ \frac{a}{2} c^2 \left(1 - \frac{c}{\cIn}\right)^2 + \frac{\kappa}{2}|\nabla c|^2 \right]
    \;,
    \label{eq:freeenergy}
\end{equation}
where $a$ sets the energy scale, $\cIn$ denotes the concentration difference between coexisting phases, and $\kappa$ penalizes gradients, implying a surface tension of $\gamma = \frac16 \cIn^2 \sqrt{\kappa \, a}$  and an interfacial width of $w = \sqrt{\kappa/a}$~\cite{weberPhysicsActiveEmulsions2019}.
The concentrations of coexisting equilibrium phases are thus $c=0$ and $c=c_\mathrm{in}$, which are distinct from the reaction equilibrium at $c=c_0$ since the reactions are active.
This imbalance drives continuous fluxes and flows in the system, which we analyze
in small systems with low velocities, implying that inertial effects are negligible.
Consequently, the velocity field $\boldsymbol{v}$ is governed by the Stokes equations,
\begin{align}
    \eta \nabla^2 \boldsymbol{v} - \nabla p &= c \, \nabla \mu
    &
    \nabla\cdot \boldsymbol{v} &= 0
    \label{eqn:stokes}
    \;,
\end{align}
describing balanced viscous stresses proportional to viscosity~$\eta$, hydrostatic pressure $p$, and equilibrium stresses.

\begin{figure}
    \centering
    \includegraphics[width=0.5\textwidth]{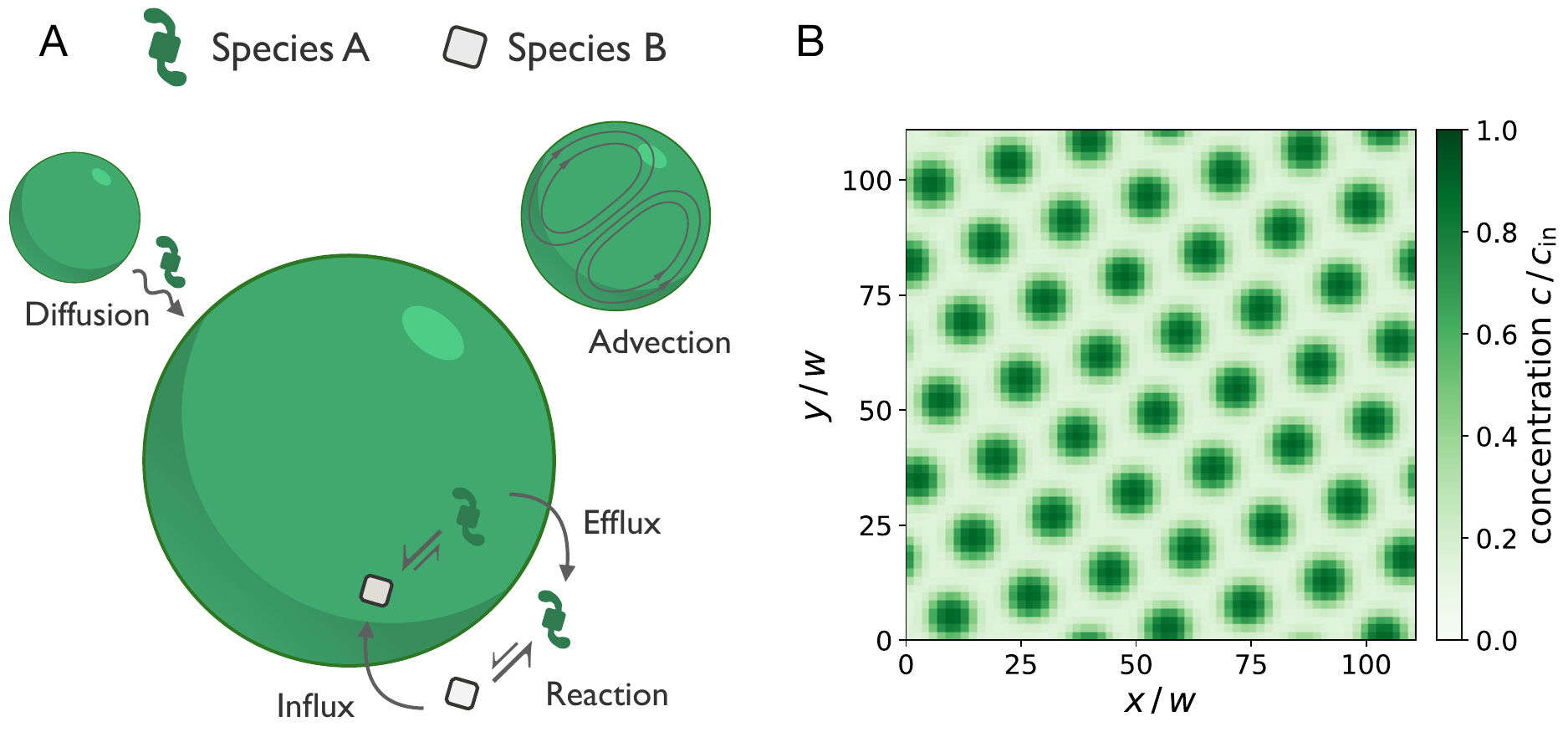}
    \caption{\textbf{Turnover causes hexagonal droplet pattern.}
    (A) Schematic of our model involving two species that inter-convert and phase separate from each other.
    (B) Simulation snapshot showing $c(\boldsymbol{r})$ at stationarity for $\Pe = 0$, $c_0 = 0.35 \, \cIn$, $k = 0.02/ \tau$, $\tau=w^2/D$, and $w = \sqrt{\kappa/a}$.
    }
    \label{fig:Schematic}
\end{figure}

We consider periodic boundary conditions since we are interested in internally created flows and neglect external influences.
To judge the relative magnitude of the terms in \Eqref{eqn:stokes}, we use the interfacial width~$w$ as a relevant length scale and estimate the associated velocity scale as $v \sim \gamma/\eta$, implying the Péclet number $\Pe = w^2 \, a \, \cIn^2/(D \, \eta)$, where $D=\Lambda a$ is the relevant diffusivity and we ignored  a proportionality factor of $\frac16$ to arrive at simple non-dimensional equations (Appendix~\ref{app:SimCH}).
To estimate realistic values for $\Pe$, we consider molecules of sizes comparable to the interfacial width~$w$, concentration $\cIn \sim w^{-3}$, and interaction energy $a \cIn \sim k_\mathrm{B}T$, where $k_\mathrm{B}T$ is the thermal energy.
Using the Stokes--Einstein relation to obtain $D\sim k_\mathrm{B}T/(6\pi\eta w)$, we find $\Pe \sim 19$.
Since $\Pe$ is not small compared to $1$, advection is likely relevant in our system.

\subsection{Chemical reactions arrest droplet coarsening}

\begin{figure}
    \centering
    \includegraphics[width=0.5\textwidth]{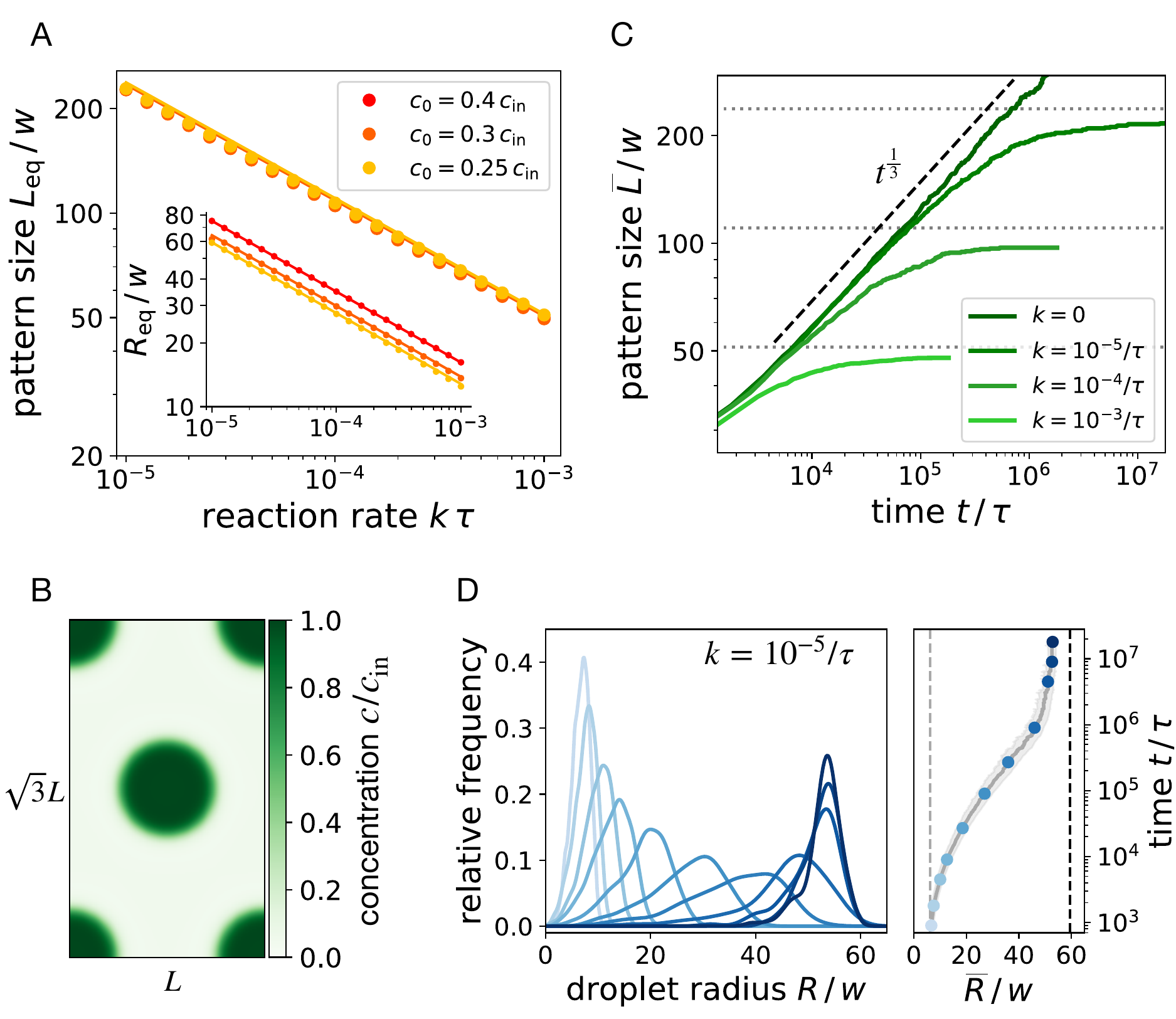}
    \caption{\textbf{Reaction rate controls stationary pattern size.}
    (A) Pattern size $L_\mathrm{eq}$, determined numerically (symbols) and predicted by  \Eqref{eq:AnalyticalLengthScale} (lines), as a function of reaction rate $k$ for various average compositions $c_0$.
    Inset shows corresponding radii.
    (B) Concentration profile $c(\boldsymbol{r})$ for $k=10^{-3}/\tau$ in a unit cell with adjustable length $L$. 
    (C) Average pattern size $\bar L$ as a function of time for various $k$, where $\bar L= 2\pi^{-1/2} (V_\mathrm{sys}/N)^{1/2}$ from droplet count $N$ and system size $V_\mathrm{sys}$.
    Numerical data are compared to $L_\mathrm{eq}$ from \Eqref{eq:AnalyticalLengthScale} (gray dotted lines) and a $t^{1/3}$ power law (black dashed line). 
    (D) Distribution of droplet radii $R$ (left) and the average radius $\bar R$ (right) for various time points for $k = 10^{-5}/\tau$.
    Numerical data are compared to the prediction from spinodal decomposition (gray dashed line) and $R_\mathrm{eq} = \sqrt{c_0/4 \cIn} \, L_\mathrm{eq}$ (black dashed line).
    (A--D)
    Model parameters are $c_0 = 0.25\,\cIn$, $\tau = w^2/D$, and $w = \sqrt{\kappa/a}$.
    }
    \label{fig:Stat_Pat_Size}
\end{figure}

\subsubsection{Stationary droplets are larger for weaker reactions} %
We start by investigating the interplay between diffusion and reactions, neglecting hydrodynamics for now ($\Pe=0$, implying $\boldsymbol{v}=\boldsymbol{0}$). %
Numerical simulations in two dimensions suggest that reactions suppress coarsening, so that droplets form a hexagonal pattern (\Figref{fig:Schematic}B).
To understand how chemical reactions  alter material fluxes to suppress coarsening, we map the active system described by \Eqref{eq:AdvReacCH} to an equivalent passive system with non-local interactions.
In particular, we rewrite the dynamics using a surrogate free energy~\cite{liuDynamicsPhaseSeparation1989,muratovTheoryDomainPatterns2002}
\begin{equation}
    \tilde{F} = F + \frac{k}{2\Lambda} \int \diff V \, \psi (c - c_0)
    \label{eq:surrogate_free_energy}
    \;,
\end{equation}
where the potential $\psi$ is governed by the Poisson equation
\begin{equation}
    \nabla^2 \psi = -(c - c_0)
    \;.
    \label{eq:Poisson}
\end{equation}
This potential captures the effect of reactions as nonlocal interactions, so that the dynamics $\partial_t c = \Lambda \nabla^2 \delta \tilde F/\delta c$ are identical to \Eqref{eq:AdvReacCH} when $\boldsymbol{v}=\boldsymbol{0}$. %
\Eqsref{eq:surrogate_free_energy}--\eqref{eq:Poisson} reveal that the reactions are analogous to an electrostatics problem, where $(c - c_0)$ acts as a charge density creating an electrostatic potential~$\psi$.
Droplets can thus be interpreted as positively charged disks, surrounded by a negative charge cloud extending over the reaction-diffusion length $\xi = \sqrt{D/k}$~\cite{zwickerPhysicsDropletRegulation2025,ziethenHeterogeneousNucleationGrowth2024}.
If the average concentration of the system reached its stationary state value $c_0$, the mathematical mapping to the equilibrium system exhibits the same dynamics as the original system given by \Eqref{eq:AdvReacCH}.
Consequently, minimizing $\tilde F$ allows us to identify stationary states of the original system.
Our numerical minimization described in Appendix~\ref{app:nummin}, reveals that the pattern size, quantified by the center-to-center distance~$L$, decreases with higher reaction rates~$k$ (\Figref{fig:Stat_Pat_Size}A), consistent with previous work~\cite{glotzerReactionControlledMorphologyPhaseSeparating1995, christensenPhaseSegregationDynamics1996, muratovTheoryDomainPatterns2002,zwickerSuppressionOstwaldRipening2015}.

We next derive an analytical estimate of the pattern size $L$ by focusing on a single droplet of radius~$R$ and approximating its environment as a spherically symmetric shell with radius $\frac12L$ (Appendix~\ref{app:statLength}).
To determine the relation between $R$ and $L$, we focus on two-dimensional systems and approximate the  concentration profile as $c(r) = \cIn\Theta(R-r)$, which describes droplets without reactions and surface tension effects.
Since the average concentration is $c_0$ in steady state, we conclude that $R/L = \sqrt{c_0/4\cIn}$, based on a spherically symmetric system.
Inserting this together with the solution of \Eqref{eq:Poisson} into \Eqref{eq:surrogate_free_energy} leads to an expression for $\tilde F$ that only depends on $L$.
Minimizing $\tilde F$ with respect to $L$ yields (Appendix~\ref{app:statLength}) %
\begin{equation}
    \frac{L_\mathrm{eq}}{w} = \frac{4}{6^{\frac13}} \!\left(\frac{\cIn}{c_0}\right)^{\frac12} \!\!\left[ \! \frac{c_0}{\cIn} - 1 - \mathrm{ln}\left( \frac{c_0}{\cIn} \right) \! \right]^{-\frac13} \!\!\!\left(\frac{w^2k}{D}\right)^{\!-\frac13} \!\!\!\!,
    \label{eq:AnalyticalLengthScale}
\end{equation}
consistent with our data (\Figref{fig:Stat_Pat_Size}A).
This expression is proportional to Eq.~(58) in ref.~\cite{muratovTheoryDomainPatterns2002}, but our prefactor approximates the data better.
In any case, material turnover leads to stationary states with controlled droplet sizes with $L_\mathrm{eq},  R_\mathrm{eq} \propto k^{-1/3}$.

\subsubsection{Chemically active droplets coarsen until their stationary size}

We next analyze the approach toward stationary state.
To analyze generic situations, we start with a slightly perturbed homogeneous state and simulate spinodal decomposition (Appendix~\ref{app:effsim}).
Once proper droplets with a narrow size distribution form, we switch to an effective simulation method to cover long timescales~\cite{kulkarniEffectiveSimulationsInteracting2023}.
The average pattern size $\bar{L}$ generally increases over time and saturates at a level that depends on $k$ (\Figref{fig:Stat_Pat_Size}C).
During the coarsening phase, we find $\bar{L}\propto t^{1/3}$ (\Figref{fig:Stat_Pat_Size}C), consistent with Lifshitz--Slyzov--Wagner theory~\cite{lifshitzKineticsPrecipitationSupersaturated1961,wagnerTheorieAlterungNiederschlaegen1961}.
Moreover, our data suggest that the droplet size distribution is similar to the expected universal one during coarsening~\cite{lifshitzKineticsPrecipitationSupersaturated1961,wagnerTheorieAlterungNiederschlaegen1961}, but it becomes much narrower close to the stationary state (\Figref{fig:Stat_Pat_Size}D)~\cite{bauermannTheoryReversedRipening2025}. 
Interestingly, the pattern size saturates below the length scale predicted by \Eqref{eq:AnalyticalLengthScale} (\Figref{fig:Stat_Pat_Size}C).

\subsubsection{Chemical reactions introduce multistability}

\begin{figure}
    \centering
    \includegraphics[width=0.5\textwidth]{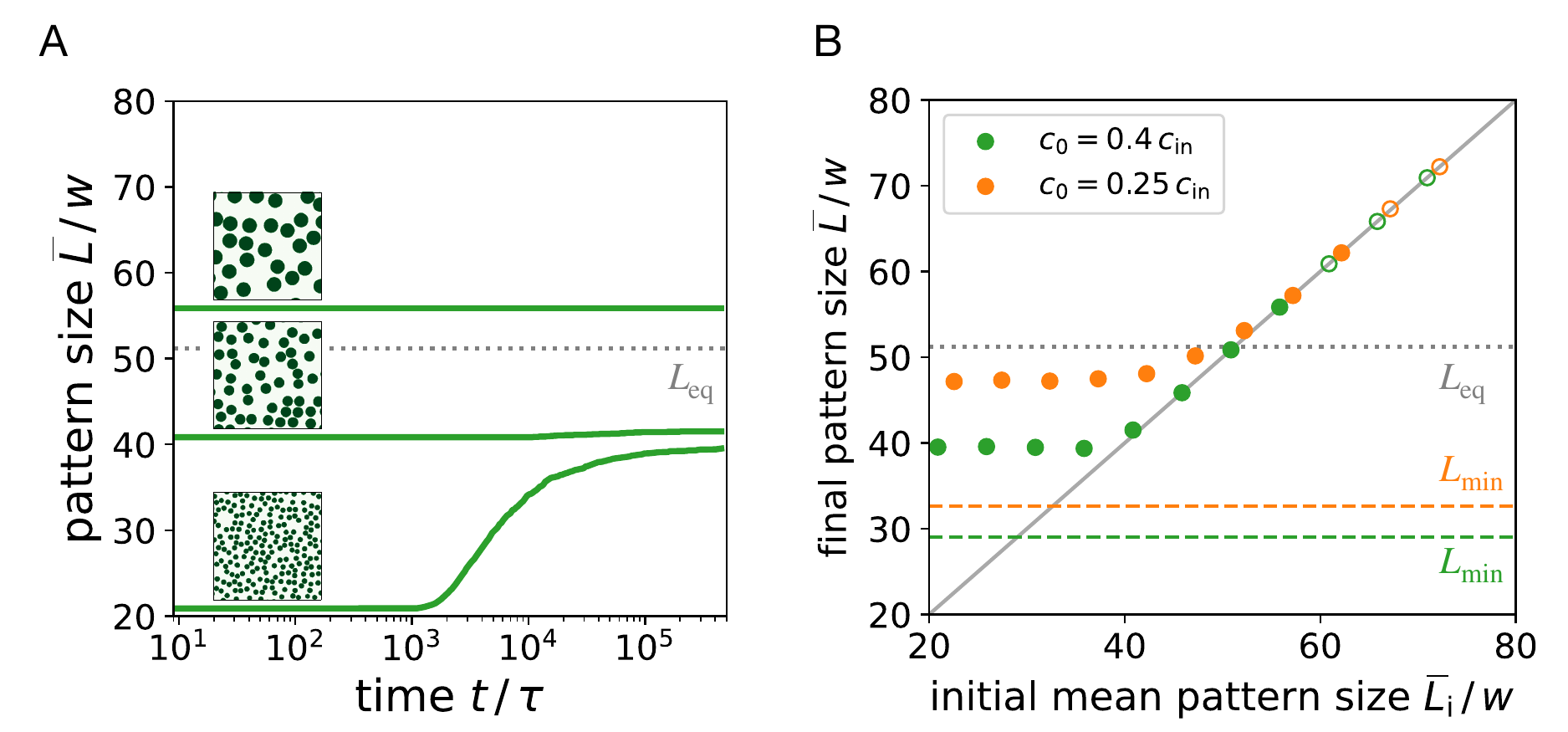}
    \caption{\textbf{Many stationary states are stable.}
    (A) Pattern size $\bar L$ as a function of time $t$ for three initial conditions with $c_0 = 0.4 \, \cIn$ and differing droplet size (see snapshots and Appendix~\ref{app:effsim}).
    (B) Final pattern size $\bar{L} = 2/\sqrt{\pi} (V_\mathrm{sys}/N)^{1/2}$ as a function of the initial pattern size $\bar{L}_\mathrm{i}$ for various $c_0$.
    Colored dashed lines indicate minimal stable size $L_\mathrm{min}$ obtained from stability analysis.
    Open circles indicate shape instabilities.
    (A--B) Gray dotted lines mark free energy minimum for $c_0 = 0.4 \, \cIn$ given by \Eqref{eq:AnalyticalLengthScale}. Model parameters are $k = 10^{-3}/\tau$, $\tau = w^2/D$, and $w = \sqrt{\kappa/a}$.
    }
    \label{fig:Multistability}
\end{figure}

To understand the discrepancy between the predicted pattern size and the final state of the simulation, we next analyze more diverse initial conditions.
Specifically, we vary the initial average droplet size, compensating the droplet count to keep the overall amount of material the same.
\Figref{fig:Multistability}A shows that coarsening occurs if the initial droplets are small, whereas large droplets remain basically unchanged.
This implies that there are many different (meta-)stable stationary states with various pattern sizes above a minimal size.
Our numerical analysis suggests that this minimal size depends on the average concentration $c_0$ (\Figref{fig:Multistability}B).

To understand the minimal size of patterns, we next analyze their stability.
Our simulations indicate that patterns coarsen by dissolving droplets, whereas coalescence is negligible in the late stage of coarsening. %
To determine the smallest droplet that can be stable, we consider a system comprising a droplet of radius $R_1$, surrounded by droplets of radii $R_2$ at distance $L$, assuming spherical symmetry (Fig.~\ref{fig:schematicStabAna}). 
The dynamics of $R_1$ are governed by $\partial_t R_1 = (j_\mathrm{in} - j_\mathrm{out})/\cIn$, where $j_\mathrm{in}$ and $j_\mathrm{out}$ are the diffusive fluxes inside and outside of the interface~\cite{weberPhysicsActiveEmulsions2019}.
These fluxes can be determined analytically assuming quasistationarity, implying that $\partial_t R_1$ can be expressed as a function of $R_1$, $R_2$, and $L$.
A stability analysis for small perturbations around $R_1=R_2=R_*$ allows us to determine the minimal stable radius, $R_\mathrm{min}$ (Appendix~\ref{app:stability}).
The corresponding predicted minimal pattern size $L_\mathrm{min}$ underestimates the smallest observed patterns (\Figref{fig:Multistability}B), but it reveals the correct trends:
Droplets can be smaller if there is more material (larger $c_0$).
The dependence of $L_\mathrm{min}$ on $c_0$ is stronger than for $L_\mathrm{eq}$, but both quantities scale as $k^{-1/3}$ (Fig.~\ref{fig:StabAna} and \cite{zwickerSuppressionOstwaldRipening2015}).
The stability analysis  predicts that patterns with droplets above the minimal size $R_\mathrm{min}$ are stable.

We next ask whether there is also a maximal pattern size $L_\mathrm{max}$.
Generally, large chemically active droplets exhibit shape instabilities~\cite{zwickerGrowthDivisionActive2017}, but these are not captured by the simulation method used in \Figref{fig:Multistability}A~\cite{kulkarniEffectiveSimulationsInteracting2023}. 
To test whether instabilities would occur, we investigate the final states of our effective simulation using detailed simulations of \Eqref{eq:AdvReacCH} (Appendix~\ref{app:shape}). %
Indeed, we find that large droplets exhibit instabilities (open symbols in \Figref{fig:Multistability}B), suggesting that such systems would evolve toward smaller droplets.

In summary, we find that chemically active emulsions can in principle exhibit a range of pattern sizes, distributed around the theoretical prediction of $L_\mathrm{eq}$.
The observed pattern size thus depends on the initial condition. 
The surrogate equilibrium model given by \Eqref{eq:surrogate_free_energy} suggests that this multistability can be interpreted as a kinetic arrest, and noise would allow the system to explore various states. 
This implies that the droplet count varies over time by nucleation, dissolution, coalescence, and splitting of droplets.
These effects are generally driven by molecular diffusion and Brownian motion of entire droplets, which depends on hydrodynamic effects.

\subsection{Hydrodynamic advection accelerates coarsening by inducing coalescence}

\begin{figure*}
    \centering
    \includegraphics[width=0.9\textwidth]{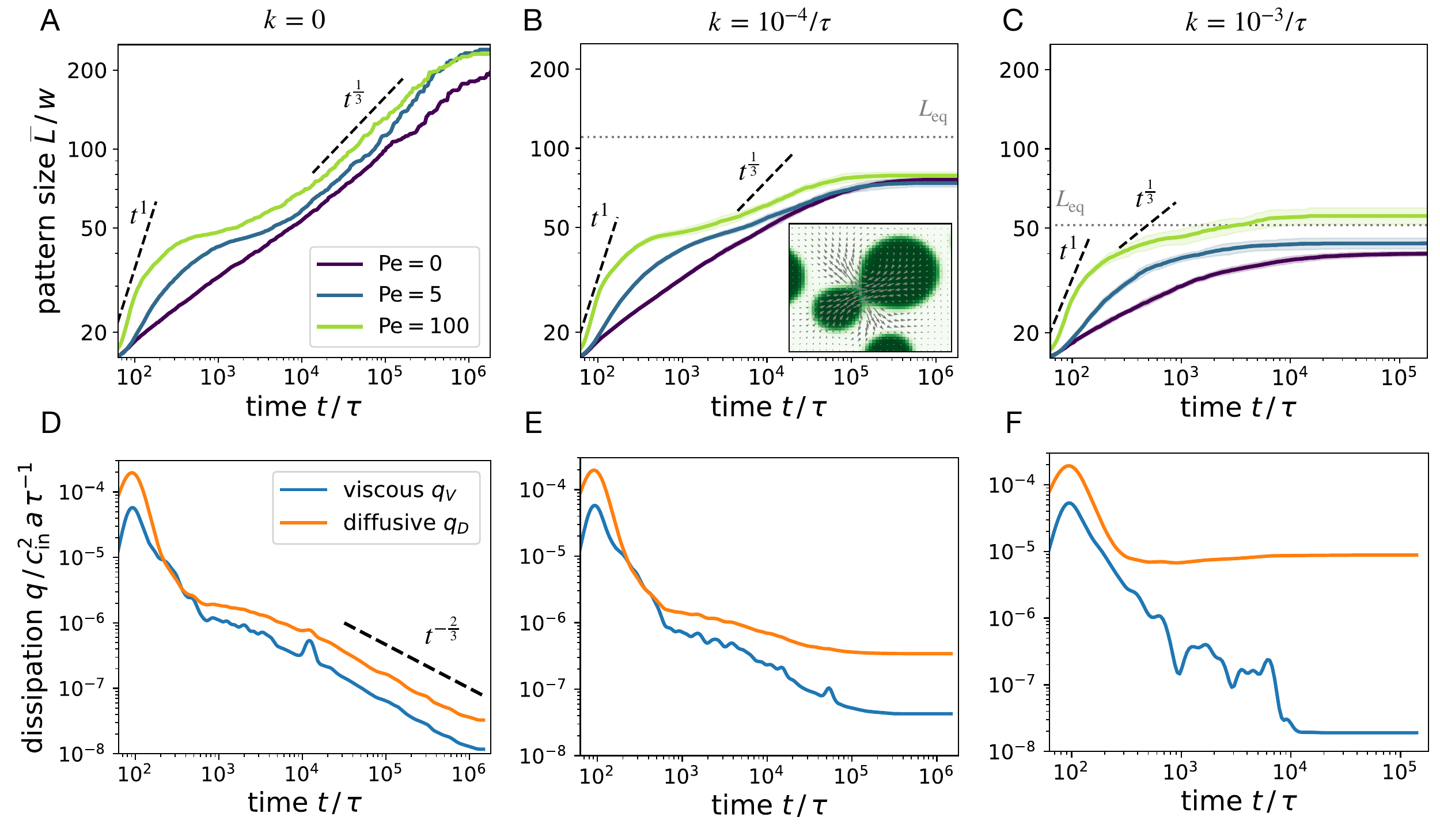}
    \caption{\textbf{Advection accelerates coarsening.}
    Pattern size $\bar L$ (shaded area indicates standard deviation, $n=20$) as a function of time $t$ for various Péclet numbers $\Pe = w^2 \, a \, \cIn^2/(D \eta)$ and reaction rates $k$ (across panels).
    Black dashed lines show power laws $\bar L \propto t$ (left) and $\bar L \propto t^{1/3}$ (right).
    Gray dotted lines show $L_\mathrm{eq}$ calculated from \Eqref{eq:AnalyticalLengthScale}.
    Inset in (B) shows the velocity field generated by two colliding droplets (gray arrows) for $k = 10^{-4}/\tau$ and $\Pe = 100$.
    (D--F) Volume averaged viscous (blue) and diffusive (orange) dissipation corresponding to (A--C) for $\Pe = 100$ (Appendix~\ref{app:dissipation}). 
    The curves were smoothed by convolution with a Gaussian kernel with a standard deviation of 2 data-points to make the plot more readable.
    Black dashed line in (D) shows a power law $\propto t^{-2/3}$.
    (A--F) Model parameters are $c_0 = 0.4\,\cIn$, $\tau = w^2/D$, and $w = \sqrt{\kappa/a}$.
    }
    \label{fig:HydrodynamicEffects}
\end{figure*}

Finally, we ask how hydrodynamics affects the coarsening of chemically active droplets by considering $\Pe>0$. 
We perform numerical simulations of Eqs.~\eqref{eq:AdvReacCH}--\eqref{eqn:stokes}, where we solve the Stokes equation using a pseudospectral method (Appendix~\ref{app:SimCH}).
To form droplets effectively, we consider the spinodal decomposition regime.
We find that hydrodynamics generally accelerates coarsening for all reaction rates (\Figref{fig:HydrodynamicEffects}), consistent with literature~\cite{nikolayevNewHydrodynamicMechanism1996,tanakaNewMechanismsDroplet1997,huoHydrodynamicEffectsPhase2003}. %
This acceleration is less pronounced for a smaller average fraction~$c_0$ (Fig.~\ref{fig:smallc0}), indicating that it relies on a large droplet density.

For passive mixtures ($k=0$, \Figref{fig:HydrodynamicEffects}A), our data at early times is consistent with the theoretically expected scaling of $\bar L \propto t$ resulting from “coalescence-induced coalescence” in the viscous regime~\cite{nikolayevNewHydrodynamicMechanism1996, tanakaNewMechanismsDroplet1997,wagnerPhaseOrderingOftwodimensional2001,gsellPhaseSeparationDynamics2022}.
In contrast, we observe a transition to Ostwald ripening at large length scales and timescales, where the mean pattern size obeys $\bar L \propto t^{1/3}$.
Interestingly, we observe a plateau region between these two regimes where $\bar L$ hardly changes, suggesting that Ostwald ripening is delayed. 
Note that the transition between the two regimes is opposite to what has been described for bicontinuous structures in equivalent systems, where diffusion dominates at small scales (implying $\bar L \propto t^{1/3}$) and advection becomes important later (leading to $\bar L\propto t$)~\cite{siggiaLateStagesSpinodal1979}.
This difference might be caused by the fact that advection is less important in our system as judged by the smaller  viscous dissipation compared to diffusive dissipation (\Figref{fig:HydrodynamicEffects}D).
However, the identical scaling of the dissipations indicates that both advection and diffusion are relevant for the entire coarsening process of passive mixtures.

When mixtures are active ($k>0$), coarsening must eventually cease since reactions limit the pattern sizes~$\bar L$. 
For weak reactions, the initial coarsening dynamics are virtually identical to the passive case, including the ballistic regime, a transient plateau, and a phase involving Ostwald ripening (\Figref{fig:HydrodynamicEffects}B).
As expected, $\bar L$ eventually reaches a stable pattern size, which is independent of $\Pe$ for weak reactions.
Note that the Ostwald ripening regime does not quite reach the expected scaling $\bar L \propto t^{1/3}$, but we hypothesize that this would be the case for even smaller reaction rates, which are difficult to simulate.
In contrast, the Ostwald ripening regime is completely absent for stronger reactions (\Figref{fig:HydrodynamicEffects}C).
Concomitantly, we observe that the final pattern size depends on $\Pe$ (and on initial condition, evident from larger standard deviation), consistent with the multistability described above.
Note that advection generally opposes shape deformations~\cite{seyboldtRoleHydrodynamicFlows2018}, suggesting larger maximal stable pattern sizes $L_\mathrm{max}$.
Indeed, we generally observe larger final pattern sizes for larger $\Pe$ if reactions are sufficiently strong (\Figref{fig:HydrodynamicEffects}C).
This is likely caused by strong advection leading to fewer and larger droplets in the initial phase, which then remain in the final state.
Taken together, advection selects the initial pattern size, essentially independent of $k$, but it does not affect the further coarsening significantly, so that the final pattern size is governed by the results shown in \Figref{fig:Multistability}B.

The relative importance of advection and diffusion can be quantified by the viscous and diffusive dissipation, respectively.
\Figref{fig:HydrodynamicEffects}D--F show that advection is relevant early, but decays with time, essentially independent of $k$ (blue lines are similar in panels D--F).
The diffusive dissipation also peaks early, and decays similarly to the viscous dissipation during droplet coarsening.
Interestingly, both dissipations are similar during the time where we observe a plateau in the pattern size (\Figref{fig:HydrodynamicEffects}A--C). %
Beyond this point, diffusive dissipation dominates, and it plateaus at a much larger value compared to the viscous dissipation.
This behavior indicates that energy induced by reactions leads to diffusive fluxes, whereas advection is negligible in the stationary state.
This might be surprising since diffusive fluxes and the velocity field are driven by gradients in chemical potential.
However, the fluid is incompressible ($\nabla\cdot \boldsymbol{v}=0$), so that chemical potential gradients get projected onto a solenoidal subspace.
Therefore, osmotic pressure gradients are balanced by gradients in hydrostatic pressure.
Accordingly, the velocity field remains small compared to diffusive fluxes for moderate advection strength (Fig.~\ref{fig:dissipation_velField}).
In contrast, large advection (large $\Pe$) causes a dynamic instability leading to spatiotemporal chaos~\cite{Datt2025}.

\section{Discussion}

We showed that a chemical conversion reaction can arrest coarsening at various length scales.
Generally, stronger conversion (larger $k$) leads to smaller length scales, although a range of scales are stable for each value of $k$.
The final length scale thus also depends on initial conditions, and particularly the number of droplets that are present.
We find that this initial droplet count depends strongly on advection, whereas advection is negligible in later stages.

Our work shows how regular patterns of droplets can be controlled by an interplay of chemical conversion and advection.
These processes are directly relevant for synthetic systems of chemically active droplets~\cite{jambon-puilletPhaseseparatedDropletsSwim2024, sastreSizeControlOscillations2024}, and they might affect biomolecular condensates in cells.
We thus speculate that cells exploit the discussed mechanism to control their condensates.
However, cellular systems are much more complex, and it is thus likely that additional processes affect condensates~\cite{zwickerPhysicsDropletRegulation2025}.
In particular, cellular flows, e.g., driven by molecular motors, could further affect coarsening~\cite{Stansell2006, Stratford2007,gubbalaDynamicSwarmsRegulate2024}, and inertia could be important on larger length scales~\cite{furtadoLatticeBoltzmannSimulations2006, Fielding2008}.
Moreover, cellular droplets exhibit significant thermal fluctuations, which would not only be relevant for nucleation~\cite{Ziethen2023}, but could particularly allow the system to transition from one metastable state to another.
In this case, we hypothesize that the most likely state exhibits a length scale close to $L_\mathrm{eq}$ given by \Eqref{eq:AnalyticalLengthScale}.
The multistability should then be interpreted as kinetic traps, potentially leading to glassy dynamics~\cite{zhangRandomIsotropicStructures2006}.

\section*{Acknowledgements}
We thank Noah Ziethen and Riccardo Rossetto for helpful discussions.
We gratefully acknowledge funding from the Max Planck Society and the European Research Council (ERC, EmulSim, 101044662).
S. K. acknowledges funding through a fellowship of the IMPRS for Physics of Biological and Complex Systems. S. K. acknowledges support by the study program “Biological Physics” of the Elitenetzwerk Bayern.

\bibliography{Bibliography,ManualBib}

\clearpage

\onecolumngrid

\appendix

\renewcommand\thefigure{S\arabic{figure}} 
\setcounter{figure}{0}

\section{Dimensionless equations and numerical details}
\label{app:SimCH}

Measuring time in units of $\tau = w^2/D$, length in units of the interface width $w = \sqrt{\kappa/a}$, and concentration in units of $c_\mathrm{in}$, we obtain the dimensionless form of Eqs. \eqref{eq:AdvReacCH}--\eqref{eqn:stokes},
\begin{subequations}
\begin{align}
    \p_t c + \boldsymbol{v}\cdot \nabla c &= \nabla^2 \mu - \tau k \left(c - \frac{c_0}{c_\mathrm{in}}\right)
    \label{eqn:pde_SI}
\\
    \nabla^2 \boldsymbol{v} - \nabla p &= \frac{w^2 \, a \, c_\mathrm{in}^2}{D \, \eta} \, c \nabla \mu
&
    \nabla \cdot \boldsymbol{v} &= 0 \; ,
    \label{eqn:stokes_SI}
\end{align}
\end{subequations}
with the dimensionless exchange chemical potential,
\begin{equation}
    \mu = c - 3c^2 + 2c^3 - \nabla^2 c
    \;.
\end{equation}
The three parameters governing the system are the dimensionless reaction rate $\tau k$, the steady state concentration $c_0/c_\mathrm{in}$, and the Péclet number $\Pe = (w^2 \, a \, c_\mathrm{in}^2)/(D \, \eta)$.

\label{app:hydro}
The Stokes equation (\Eqref{eqn:stokes} in main text, \Eqref{eqn:stokes_SI} above) determines the velocity field $\boldsymbol v$ for each density field~$c$.
We thus solve for the velocity field $\boldsymbol{v}$ in every time step by applying the Oseen propagator to the forcing $\boldsymbol{f} = - c \nabla \mu$ in Fourier space. 
We derive the propagator for the forced incompressible Stokes equation
\begin{align}
    \eta \nabla^2 \boldsymbol{v} &= \nabla p - \boldsymbol{f}
    &
    \nabla \cdot \boldsymbol{v} &= 0 \; ,
\end{align}
by taking the divergence and using $\nabla \cdot \boldsymbol{v} = 0$. We get
\begin{equation}
    \nabla^2 p = \nabla \cdot \boldsymbol{f}
    \;.
\end{equation}
Fourier transforming both equations leads to
\begin{align}
    \hat{\boldsymbol{v}} &= - \frac{i\boldsymbol{k} \hat{p} - \hat{\boldsymbol{f}}}{\eta \, k^2}
& \text{and} &&
    \hat{p} &= - \frac{i\boldsymbol{k} \cdot \hat{\boldsymbol{f}}}{k^2}
\end{align}
Plugging the second expression into the first, we end up with
\begin{equation}
    \hat{\boldsymbol{v}} = \frac{\hat{\boldsymbol{f}}}{\eta \, k^2} - \frac{\boldsymbol{k} (\boldsymbol{k} \cdot \hat{\boldsymbol{f}})}{\eta \, k^4} = \mathbf{P}^\mathrm{0} \cdot \hat{\boldsymbol{f}}
    \;,
\end{equation}
where 
\begin{equation}
    \mathbf{P}^\mathrm{0} = \frac{1}{\eta} \left( \frac{\mathbf{I}}{k^2} - \frac{\boldsymbol{k} \boldsymbol{k}}{k^4}\right)
\end{equation}
is the Oseen propagator, which essentially projects the force to a divergence free subspace.
Here, $\mathbf{I}$ is the identity matrix.

Eqs. \eqref{eqn:pde_SI}-\eqref{eqn:stokes_SI} are solved in a two-dimensional periodic domain.
To solve the dynamics of $c$ given by \Eqref{eqn:pde_SI} numerically, we use a finite difference method implemented in the \href{https://github.com/zwicker-group/py-pde}{\textit{py-pde}} Python package~\cite{Zwicker2020}.
During each time step, we use the projection method presented above using a pseudo-spectral method to determine $\boldsymbol{v}$.
We use a Cartesian grid of size $(128\times128)\sqrt{3}w$ for $k =0$ and $k = 10^{-3}/\tau$, and $(256\times256)\sqrt{3}w$ for $k = 10^{-4}/\tau$ with resolution $\Delta x = \sqrt{3} w$.
In our simulation we choose $w = \frac{1}{\sqrt{3}}$, $\tau = \frac19$, an initial time step $\Delta t = 0.045\tau$, and we use adaptive time stepping.
We initialize the system with an average concentration $\bar{c}_\mathrm{init} = c_0$ with added noise drawn from a uniform distribution.
For each Péclet number $\Pe$ we run 20 simulations to report means and standard deviations.

\begin{figure}
    \centering
    \includegraphics[width=0.8\textwidth]{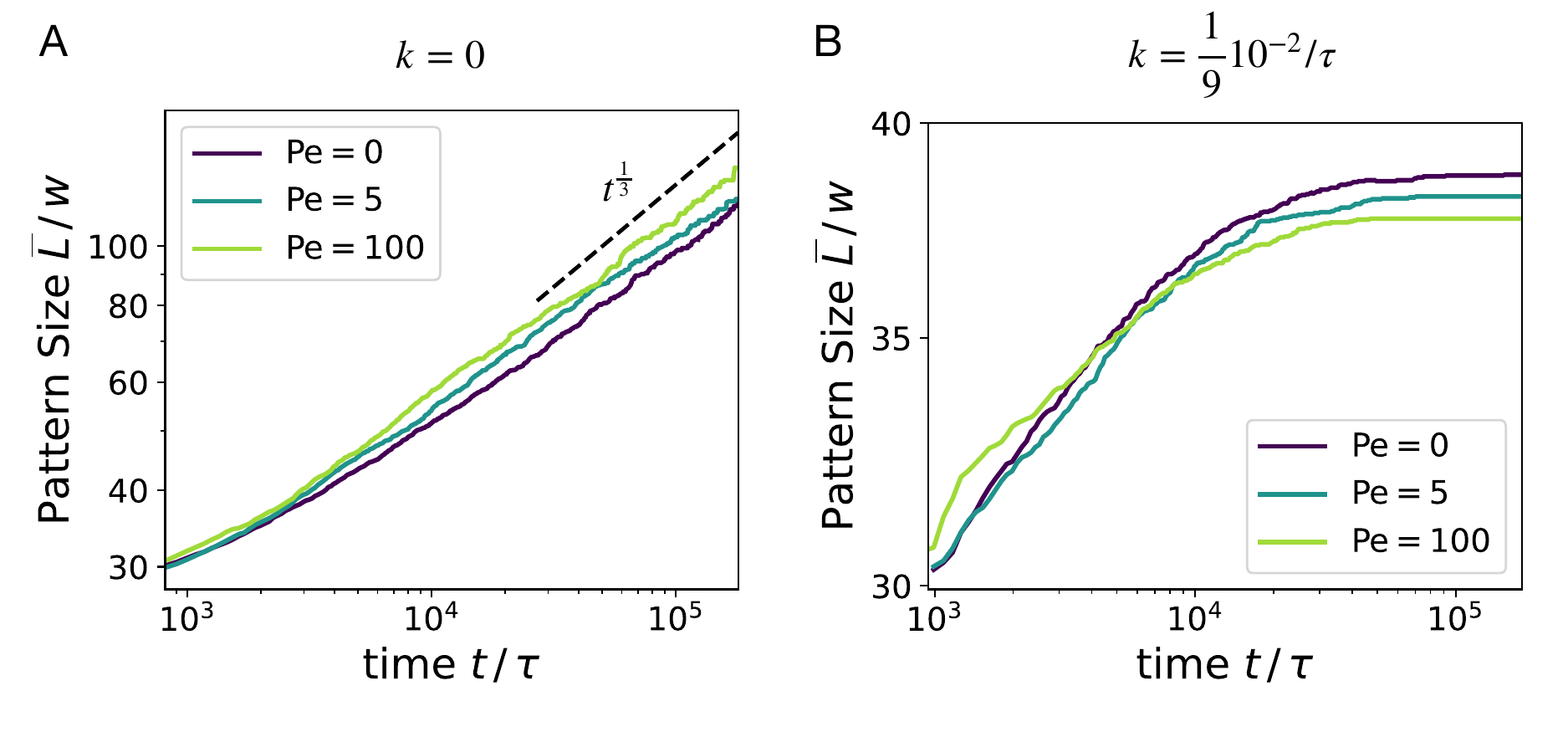}
    \caption{Pattern size $\bar L$ (for $n=20$ simulations per line) as a function of time $t$ for various Péclet numbers $\Pe = w^2 \, a \, c_\mathrm{in}^2/(D \eta)$ and reaction rates $k$ (across panels).
    Black dashed line shows power laws $\bar L \propto t^{1/3}$ (A).
    (A--B) Model parameters are $c_0 = 0.25\,c_\mathrm{in}$, $\tau = w^2/D$, and $w = \sqrt{\kappa/a}$.}
    \label{fig:smallc0}
\end{figure}

\subsection{Effective simulations}
\label{app:effsim}

We use an effective droplet simulation method~\cite{kulkarniEffectiveSimulationsInteracting2023} to run simulations in 2D with low reaction rates, where the system volume needs to be large to get proper statistics, and the simulation time increases with $k^{-1}$, 

For the simulations shown in \Figref{fig:Stat_Pat_Size}, the initial conditions were generated by performing a simulation of \Eqref{eq:AdvReacCH} with $\Pe = 0$ for a noisy initial concentration $\bar{c}_\mathrm{init} = c_0$ for a time $T = 1350 \, \tau$ for $\tau k = 10^{-3}$ and $T = 900 \, \tau$ for larger values of $\tau k$.
Subsequently, droplets are identified with the \href{https://github.com/zwicker-group/py-droplets}{\textit{py-droplets}} Python package and the concentration of the background field outside the droplets is calculated according to mass conservation. The calculated value fits well when compared to the outside concentration field in the Cahn--Hilliard simulation.

\begin{figure}
    \centering
    \includegraphics[width=0.4\textwidth]{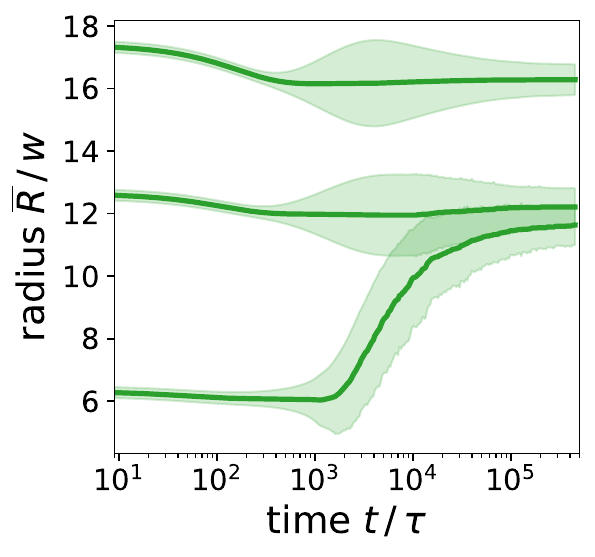}
    \caption{
        Mean radius of droplet emulsions over time corresponding to curves in \Figref{fig:Multistability}A in the main text. Shaded areas depict the standard deviation of droplet radii.
        Parameters of the effective simulation are $k = 10^{-3}/\tau$, $c_0 = 0.4 c_\mathrm{in}$, $\tau = w^2/D$, and $w = \sqrt{\kappa/a}$.
    }
    \label{fig:R_Multistab}
\end{figure}

For simulations shown in \Figref{fig:Multistability}, the droplets where initialized sequentially in space according to a uniform distribution and their radii drawn from a normal distribution $\mathcal{N}(\bar{R}_0, \sigma_R^2)$.
If an overlap between a potential droplet and an already existing droplet is detected, the droplet is discarded, and a new droplet is drawn from the distribution.
The mean radius of the distribution is calculated from the given initial mean pattern size $\bar{L}_\mathrm{i}$ according to $\bar{R}_0 = \sqrt{\frac{c_0}{4 \, c_\mathrm{in}}} L_\mathrm{i}$.
The standard deviation is set to $\sigma_R = 0.1 \sqrt{3} \, w$.
We chose the concentration field outside the droplets as $\bar{c}_\mathrm{out}^\mathrm{eq} = \frac{2\gamma}{a \, c_\mathrm{in}^2 \, \bar{R}_0}$ (equilibrium concentration outside a droplet of radius $\bar{R}_0$) to avoid collective growth/shrinking of droplets.
To guarantee a mean concentration of $c_\mathrm{init} = c_0$ the number of droplets is calculated to be 
\begin{equation}
    N = \mathrm{Int}\left[\frac{V_\mathrm{Sys}}{\pi \bar{R}_0^2} \frac{c_0 - \bar{c}_\mathrm{out}^\mathrm{eq}}{c_\mathrm{in} - \bar{c}_\mathrm{out}^\mathrm{eq}} \right]
    \;.
\end{equation}
\Figref{fig:R_Multistab} shows that the mean radius first decreases, implying droplets loose mass to their surroundings.
Afterwards, the coarsening process starts, leading to a broad distribution of radii.

\section{Surrogate equilibrium model}

This section describes our strategy to analyze the stationary state of the dynamical equation
\begin{equation}
    \p_t c = \Lambda \nabla^2 \mu - k(c - c_0)
    \;.
    \label{eqn:dynamics_SI}
\end{equation}

\subsection{Derivation of free energy}
To determine stationary states of \Eqref{eqn:dynamics_SI}, we exploit an analogy to electrostatics and introduce the potential $\psi$,
\begin{equation}
    \nabla^2 \psi = -(c - c_0)\;,
    \label{eq:Potential}
\end{equation}
so the dynamical equation becomes
\begin{equation}
    \p_t c = \Lambda \nabla^2 \left( \mu + \frac{k}{\Lambda} \psi\right)\;.
\end{equation}
Integrating over the equation, we get the additional constraint
\begin{align}
    \int_\Omega \diff V \p_t c &= \Lambda \int_\Omega \diff V \nabla^2 \mu - k \int_\Omega \diff V(c - c_0) \notag\\ 
     V \p_t \bar{c} &= \Lambda \cdot 0- k V  (\bar{c} - c_0) \notag\\ 
    \p_t \bar{c} &= - k  (\bar{c} - c_0).
\end{align}
where $V=\int_\Omega\diff V$ is the system's volume.
The average concentration in the system thus decreases exponentially to $c_0$.
Therefore, we get the constraint 
\begin{equation}
    \int_\Omega \diff V \, c = c_0 V
    \label{eq:elstatconstraint}
\end{equation}
on the stationary state.
We can then introduce an extended free energy 
\begin{equation}
    \tilde{F} = F + \frac{k}{2\Lambda} \int \diff V \, \psi (c - c_0)
\end{equation}
with
\begin{equation}
    \frac{\delta\tilde{F}}{\delta c} = \mu + \frac{k}{\Lambda} \psi
\end{equation}
since
\begin{align}
    \delta\tilde{F} &= \delta F + \frac{k}{2\Lambda} \int \diff V \left[\delta\psi \, (c - c_0) + \psi \, \delta c\right] \notag \\
    &= \delta F + \frac{k}{2\Lambda} \int \diff V \left[-\delta\psi \, \p_i^2 \psi + \psi \, \delta c\right] \notag \\
    &\overset{2\times \mathrm{P.I.}}{=} \delta F + \frac{k}{2\Lambda} \int \diff V \left[-\delta(\p_i^2 \psi) \,  \psi + \psi \, \delta c\right] \notag \\
    &= \delta F + \frac{k}{2\Lambda} \int \diff V \left[\delta c \,  \psi + \psi \, \delta c\right] = \delta F + \frac{k}{2\Lambda} \int \diff V (2\psi) \delta c
    \;.
\end{align}
Inserting \Eqref{eq:Potential} and integrating by parts results in
\begin{equation}
    \tilde{F} = F + \frac{k}{2\Lambda} \int \diff V \, |\nabla \psi|^2
    \label{eqn:surrogate_SI}
    \;.
\end{equation}
By construction, minima of $\tilde F$ correspond to stationary state of \Eqref{eqn:dynamics_SI}.

\subsection{Numerical minimization}
\label{app:nummin}

We first discuss how we minimize the surrogate free energy (\Eqref{eq:surrogate_free_energy} in the main text, \Eqref{eqn:surrogate_SI} above) numerically.
We perform the minimization with a variable cell method, where we not only optimize the profiles in a periodic computational box with self-consistent iterations, but also optimize the periods $L_i$ in each direction at the same time.

We first embed the Poisson equation and the condition that $\int \mathrm{d}V (c - c_0) = 0$ at the stationary state into the surrogate free energy,
\begin{equation}
    \hat{F}[c,\psi;\xi] = \int \mathrm{d}V \left[ \frac{a}{2} c^2 \left(1 - \frac{c}{c_\mathrm{in}}\right)^2 + \frac{\kappa}{2}|\nabla c|^2 + \frac{k}{\Lambda} \psi (c - c_0) - \frac{k}{2\Lambda} |\nabla \psi|^2 + \xi (c - c_0) \right] 
    \;,
\end{equation}
where $\hat{F}$ is a functional of both composition $c$ and the potential $\psi$, and a function of the Lagrange multiplier $\xi$.
With this form, the Poisson equation results from minimizing $\hat{F}$ with respect to the potential $\psi$, whereas the condition $\int \mathrm{d}V (c - c_0) = 0$ results from minimizing $\hat{F}$ with respect to $\xi$.
Note that $\hat{F}$ recovers the original surrogate free energy when we insert these two extra equations, thus the minima of $\hat{F}$ are identical to those of $\tilde{F}$.
Therefore, $\hat{F}$ is the single functional that we can minimize without any constraint, which is more convenient for numerical schemes.
To find the optimal stationary profile and periods, we further decouple the periods $L_i$ in the two spatial directions $i=1,2$ from the field variables.
The free energy density thus read
\begin{equation}
    \hat{f}[\hat{c},\hat{\psi};\xi, \{L_i\}] = \int \mathrm{d}\hat{V} \left[ \frac{a}{2} \hat{c}^2 \left(1 - \frac{\hat{c}}{c_\mathrm{in}}\right)^2 + \frac{\kappa}{2} \sum_i \frac{(\partial_i \hat{c})^2}{L_i^2} + \frac{k}{\Lambda} \hat{\psi} (\hat{c} - c_0) - \frac{k}{2\Lambda} \sum_i \frac{(\partial_i \hat{\psi})^2}{L_i^2} + \xi (\hat{c} - c_0) \right] 
    \;,
\end{equation}
where $\hat{c}$, $\hat{\psi}$, and the integral are all defined on a unit square.
Minimizing $\hat{f}$ with respect to all variables, we obtain the self-consistent equations for the stationary state,
\begin{subequations}
    \begin{align}
        \frac{\delta \hat{f}}{\delta c} &= a \hat{c} \left(1 - \frac{\hat{c}}{c_\mathrm{in}}\right)\left(1- 2\frac{\hat{c}}{c_\mathrm{in}}\right) - \kappa \sum_i \frac{\partial_i^2 \hat{c}}{L_i^2} + \frac{k}{\Lambda} \hat{\psi} + \xi = 0 \\
        \frac{\delta \hat{f}}{\delta \psi} &= \frac{k}{\Lambda} (\hat{c} - c_0) + \frac{k}{\Lambda} \sum_i \frac{\partial_i^2 \hat{\psi}}{L_i^2} = 0\\
        \frac{\partial \hat{f}}{\partial \xi} &= \int \mathrm{d}\hat{V} (\hat{c} - c_0) = 0 \\
        \frac{\partial \hat{f}}{\partial L_i} &= - \kappa \frac{(\partial_i \hat{c})^2}{L_i^3} + \frac{k}{\Lambda} \frac{(\partial_i \hat{\psi})^2}{L_i^3} = 0
        \;.
        \label{eq:sc}
    \end{align}
\end{subequations}
Note that the last equation above indicates that the interfacial energy and the electrostatic energy reach balance on each direction in the stationary state.
We solve \Eqref{eq:sc} by the following iteration scheme,
\begin{subequations}
    \begin{align}
        \sum_i \frac{\partial_i^2 \hat{\psi}}{L_i^2} &= - (\hat{c} - c_0) \\
        \Delta \hat{c} &= - a \hat{c} \left(1 - \frac{\hat{c}}{c_\mathrm{in}}\right)\left(1- 2\frac{\hat{c}}{c_\mathrm{in}}\right) + \kappa \sum_i \frac{\partial_i^2 \hat{c}}{L_i^2} - \frac{k}{\Lambda} \hat{\psi} \\
        \xi &= \int \mathrm{d}\hat{V} \Delta c \\
        \hat{c}^* &= \hat{c} + \alpha (\Delta \hat{c} - \xi) \\
        L_i^* &= L_i + \beta \left(\kappa \frac{(\partial_i \hat{c})^2}{L_i^3} - \frac{k}{\Lambda} \frac{(\partial_i \hat{\psi})^2}{L_i^3}\right)
        \;.
        \label{eq:numerics}
    \end{align}
\end{subequations}
Here $\hat{c}^*$ and $L_i^*$ are the new variables used for the next iteration.
The parameters $\alpha$ and $\beta$ are two empirical acceptance parameters to improve the stability of the iteration.
We choose $\alpha = 0.02$ and $\beta = 1$ in our numerics.
The iterations are initialized from roughly hexagonally-packed droplets in 2D.
After minimization, the periods converge to $L_2 / L_1 = \sqrt{3}$, as shown in \Figref{fig:Stat_Pat_Size}B in the main text.

\subsection{Analytical minimization}
\label{app:statLength}

We now discuss how we minimize the surrogate free energy (\Eqref{eq:surrogate_free_energy} in the main text, \Eqref{eqn:surrogate_SI} above) analytically.
To do this, we consider a system of droplets arranged in an hexagonal lattice where the centers of droplets are spaced by distance $L$.
To simply calculations, we approximate the hexagonal lattice by a circular region around each droplet with radius $L/2$.
We assume the interface thickness is much smaller than $L$ and the radius of the droplet $R$.
Furthermore, the concentrations inside and outside the droplet are assumed as homogeneous, with $c_\mathrm{out} = 0$. 
This assumption is justified if $c_\mathrm{out} \, l_c / R$ is small and $L \ll \xi$ where $\xi = \sqrt{D/k}$ is the reaction-diffusion length scale.
However, the assumption still works very well in the case $L \lesssim \xi$.
From the constraint \eqref{eq:elstatconstraint}, we get a relation between $L$ and $R$ in our circular domain 
\begin{align}
    &\pi \left(\frac{L}{2}\right)^2 \cdot c_0 = \pi R^2 \cdot c_\mathrm{in}
    &\Rightarrow&&
     R = \sqrt{\frac{c_0}{4 c_\mathrm{in}}} \, L
    \label{eq:RLrelation2}
\end{align}
We can solve for the potentials inside and outside the droplets in a radial symmetric domain of radius $L/2$,
\begin{align}
    \p_r^2 \psi_{\mathrm{in}/\mathrm{out}} + \frac{1}{r} \p_r \psi_{\mathrm{in}/\mathrm{out}} &= -(c_{\mathrm{in}/\mathrm{out}} - c_0)
    &
    \p_r \psi_{\mathrm{in}}|_{r=0} &= 0
    &
    \p_r \psi_{\mathrm{out}}|_{r=L/2} &= 0
    \;.
\end{align}
This gives
\begin{align}
    \p_r \psi_{\mathrm{in}} &= \frac{c_0 - c_\mathrm{in}}{2} r
    &\text{and}&&
    \p_r \psi_{\mathrm{out}} &= \frac{c_0 - c_\mathrm{out}}{2} \left[ r - \frac{(L/2)^2}{r} \right]
    \;.
\end{align}
Because we do not explicitly describe the interface, we need to add an extra term $\gamma 2 \pi R$ to the free energy to account for interface tension.
Using the Ginzburg-Landau model
\begin{equation}
    f_0(c) = \frac{a}{2}c^2\left(1 - \frac{c}{c_\mathrm{in}}\right)^2,
\end{equation}
the surface tension is $\gamma = \sqrt{\kappa a} \, c_\mathrm{in}^2/6$~\cite{weberPhysicsActiveEmulsions2019}.

Evaluating the extended free energy inside the circular domain we get
\begin{align}
    \tilde{F} = \ &f_0(c_\mathrm{in}) \pi R^2 + f_0(c_\mathrm{out}) (\pi (L/2)^2 - \pi R^2) + \gamma 2 \pi R + \frac{\pi k}{\Lambda} \left[ \int_0^R dr \, r |\p_r \psi_\mathrm{in}|^2 + \int_R^{L/2} dr \, r |\p_r \psi_\mathrm{out}|^2 \right].
\end{align}
Using that $f_0(c_\mathrm{in}) = f_0(c_\mathrm{out})$ and dividing by the volume of the circular domain, we get the average free energy density
\begin{align}
    \tilde{f} = f_0(c_\mathrm{in}) + 8\gamma  \frac{R}{L^2} + \frac{4 k}{\Lambda L^2} \left[ \int_0^R dr \, r |\p_r \psi_\mathrm{in}|^2 + \int_R^{L/2} dr \, r |\p_r \psi_\mathrm{out}|^2 \right].
\end{align}
Using \Eqref{eq:RLrelation2} and minimizing with respect to $L$ leads to 
\begin{equation}
    L = 4  \, c_0^{-1/2} \ c_\mathrm{in}^{-1/6} \ \left[ \frac{c_0}{c_\mathrm{in}} - 1 - \mathrm{ln}\left( \frac{c_0}{c_\mathrm{in}} \right)\right]^{-1/3} \ \gamma^{1/3} \ \left(\frac{k}{\Lambda}\right)^{-1/3}.
    \label{eq:AnalyticalLengthScale2}
\end{equation}
Using the definition of $\gamma$ and $D = \Lambda a$, we get
\begin{equation}
    \frac{L}{w} = \frac{4}{6^{1/3}}  \, \left(\frac{c_0}{c_\mathrm{in}}\right)^{-1/2} \ \left[ \frac{c_0}{c_\mathrm{in}} - 1 - \mathrm{ln}\left( \frac{c_0}{c_\mathrm{in}} \right)\right]^{-1/3} \ \left(\frac{w^2 \, k}{D}\right)^{-1/3},
    \label{eq:AnalyticalLengthScale2Dimless}
\end{equation}
which is the result given in the main text.

\section{Minimally stable droplet size from linear stability analysis}
\label{app:stability}

\begin{figure}
    \centering
    \includegraphics[width=0.7\textwidth]{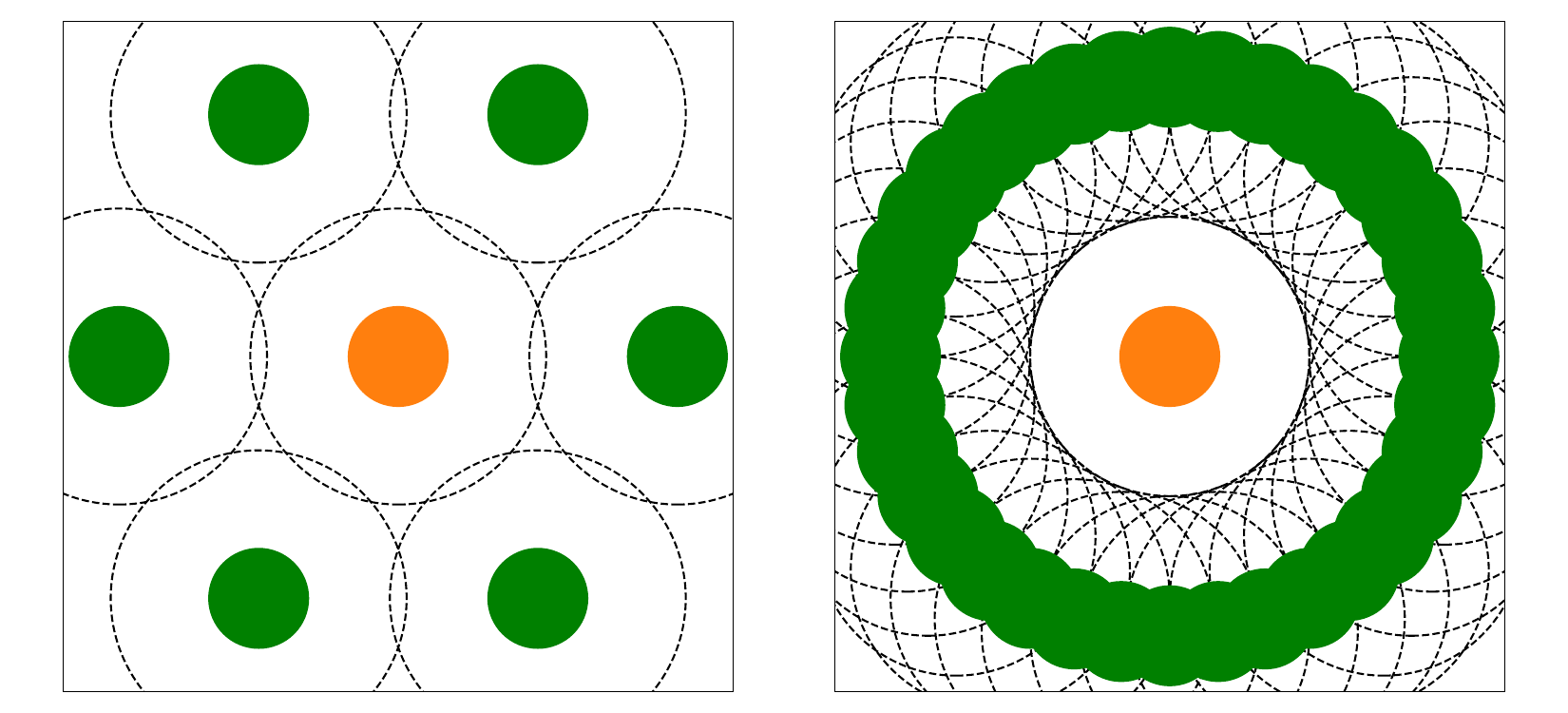}
    \caption{Schematic depiction of droplets in a hexagonal lattice (left) and the simplification to a spherically symmetric domain (right).}
    \label{fig:schematicStabAna}
\end{figure}

\begin{figure}
    \centering
    \includegraphics[width=0.7\textwidth]{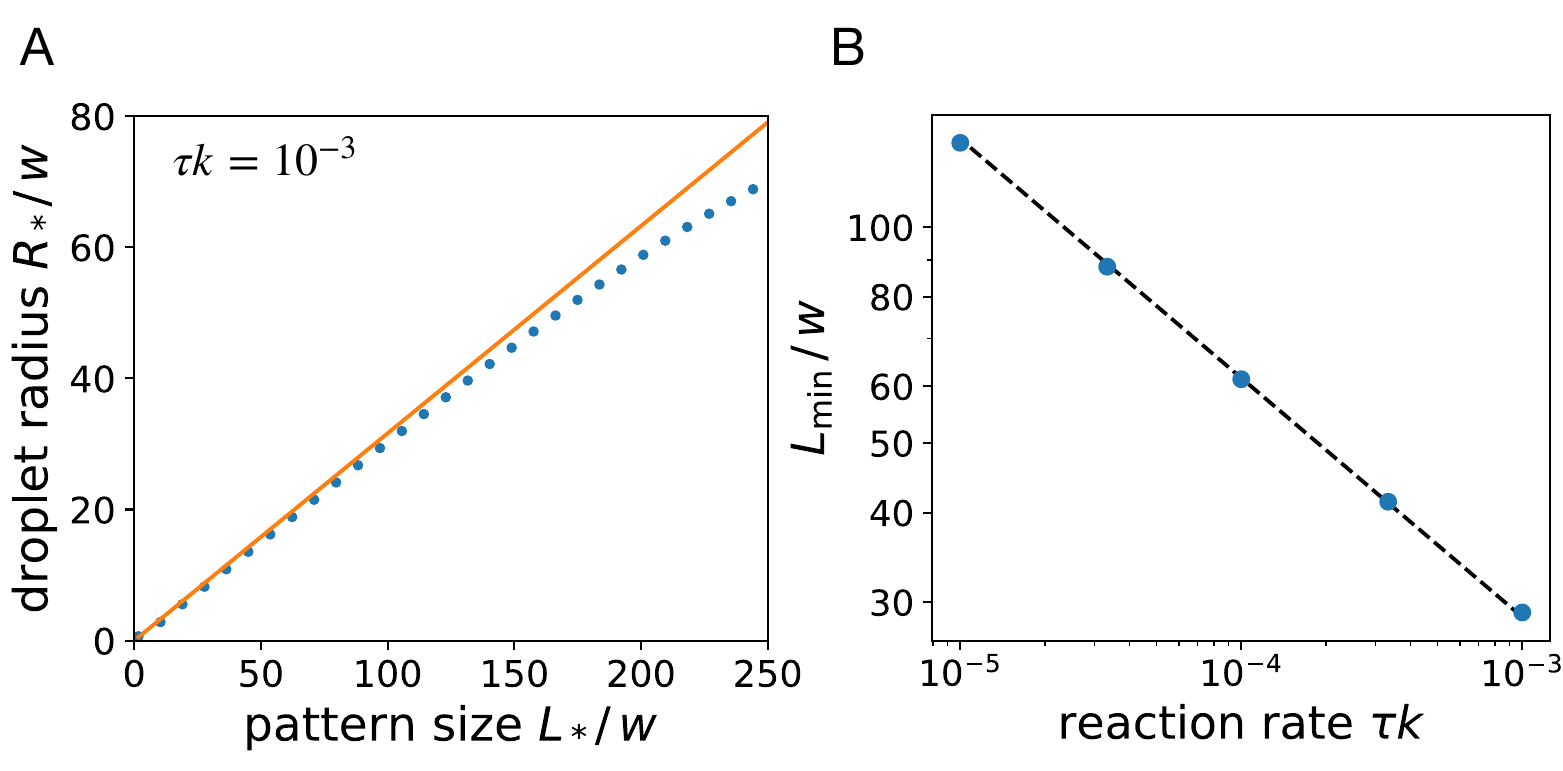}
    \caption{(A) Numerical results for $g_0(R_*,L_*) = 0$ (blue dots) compared to $R_* = \sqrt{c_0/4 c_\mathrm{in}} \, L_*$ (orange line). Parameters: $c_0 = 0.4 \, c_\mathrm{in}$, $k = 10^{-3}/\tau$. (B) Numerical results for $g_1(L_\mathrm{min}) = 0$, where $R_\mathrm{min} = \sqrt{c_0/4 c_\mathrm{in}} \, L_\mathrm{min}$ was used to write the function only in terms of $L_\mathrm{min}$. The black dashed line shows a $k^{-1/3}$ power law for comparison. Parameters: $c_0 = 0.4 \, c_\mathrm{in}$.}
    \label{fig:StabAna}
\end{figure}

We here determine the minimal stable droplet size in a hexagonal configuration.
To simply the configuration, we approximate the hexagonal configuration by a spherically symmetric one, where we imagine a system of one droplet surrounded by an infinite number of droplets (or continuous annulus) at distance $L$ (\Figref{fig:schematicStabAna}).
The inner droplet has radius $R_1$ while the outer droplets have radius $R_2$.
We can then analyze the dynamics of these two droplets in this more symmetric configuration.
From the linearized 2D reaction-diffusion equation inside and outside the droplet,
\begin{equation}
    D \left(\p_r^2 c_\mathrm{in/out} + \frac{1}{r} \p_r c_\mathrm{in/out}\right) = k (c_\mathrm{in/out} - c_0)
    \;,
\end{equation}
we get
\begin{equation}
    c_\mathrm{in/out} = c_0 + A_\mathrm{in/out} I_0\left(\frac{r}{\xi}\right) + B_\mathrm{in/out} K_0\left(\frac{r}{\xi}\right)
    \;,
\end{equation}
where $\xi = \sqrt{D/k}$ is the reaction diffusion length, and we have the boundary conditions
\begin{subequations}
\begin{align}
    \p_r c_\mathrm{in}|_{r=0} &= 0, &
    c_\mathrm{in}|_{r=R_1} &= c_\mathrm{in}^0, \\
    c_\mathrm{out}|_{r=R_1} &= c_\mathrm{out}^0\left(1 + \frac{l_\gamma}{R_1}\right), &
    c_\mathrm{out}|_{r=L - R_2} &= c_\mathrm{out}^0\left(1 + \frac{l_\gamma}{R_2}\right)\;.
\end{align}
\end{subequations}
Hence,
\begin{subequations}
\begin{align}
    c_\mathrm{in} =& \ c_0 + (c_\mathrm{in}^0 - c_0) \frac{I_0(r/\xi)}{I_0(R_1/\xi)}, \\
    c_\mathrm{out} =& \ c_0 + \frac{\left(c_\mathrm{out}^0\left(1 + l_\gamma/R_2\right) - c_0\right) K_0(R_1/\xi) - \left(c_\mathrm{out}^0\left(1 + l_\gamma/R_1\right) - c_0\right) K_0((L - R_2)/\xi)}{K_0(R_1/\xi) \, I_0((L-R_2)/\xi) - I_0(R_1/\xi) \, K_0((L-R_2)/\xi)} \ I_0\left(\frac{r}{\xi}\right) \notag \\
    &+ \frac{\left(c_\mathrm{out}^0\left(1 + l_\gamma/R_2\right) - c_0\right) I_0(R_1/\xi) - \left(c_\mathrm{out}^0\left(1 + l_\gamma/R_1\right) - c_0\right) I_0((L - R_2)/\xi)}{K_0(R_1/\xi) \, I_0((L-R_2)/\xi) + I_0(R_1/\xi) \, K_0((L-R_2)/\xi)} \ K_0\left(\frac{r}{\xi}\right)\;.
\end{align}
\end{subequations}
We can then calculate the fluxes right inside and outside the interface of the center droplet,
\begin{equation}
    j_\mathrm{in/out} = - D \, \p_r c_\mathrm{in/out}|_{r=R_1}
    \;.
\end{equation}
From this, we can again look at the rate of change of the center droplet's volume,
\begin{equation}
   2\pi R_1 \frac{\mathrm{d}R_1}{\mathrm{d}t} = \frac{\left( J_\mathrm{in} - J_\mathrm{out} \right)}{c_\mathrm{in}^0 - c_\mathrm{out}^0}
   \;,
   \label{eq:VolumeChange2DReacDiff}
\end{equation}
where
\begin{equation}
    J_\mathrm{in/out} = 2\pi R_1 j_\mathrm{in/out}
\end{equation}
is the total flux through the surface right inside/right outside the droplet.
For $R_1 \ll \xi$, using $I_1(x)/I_0(x) \approx x/2$, 
\begin{equation}
    J_\mathrm{in} \approx \pi \frac{R_1^2}{\xi^2} \, D (c_0 - c_\mathrm{in}^0) = V_\mathrm{d} \, k (c_0 - c_\mathrm{in}^0)
    \;.
\end{equation}
For the stability analysis of the emulsion, we expand around the stable radius of the emulsion, $R_1 = R_2 = R_*$, and perturb the radius of the inner droplet by $\delta$
\begin{equation}
    \frac{\mathrm{d} \delta}{\mathrm{d} t} = g_0(R_*,L_*) + g_1(R_*,L_*) \, \delta \; .
\end{equation}
The stationary state is governed by $g_0(R_*,L_*) = 0$ and gives a relation between $R_*$ and $L_*$.
In contrast, the sign of $g_1(R_*,L_*)$ determines if a certain pair $(R_*,L_*)$ is linearly stable to perturbations in the radius of the inner droplet. The solution to $g_0(R_*,L_*) = 0$ can be approximated by $R_* = \sqrt{c_0/4 c_\mathrm{in}} \, L_*$ (\Figref{fig:StabAna}A).
Plugging this into $g_1$ we can solve $g_1(L_\mathrm{min}) = 0$ to get the stability boundary of the emulsion (\Figref{fig:StabAna}B).

\section{Maximally stable droplet size due to shape instabilities}
\label{app:shape}
\begin{figure}
    \centering
    \includegraphics[width=0.7\textwidth]{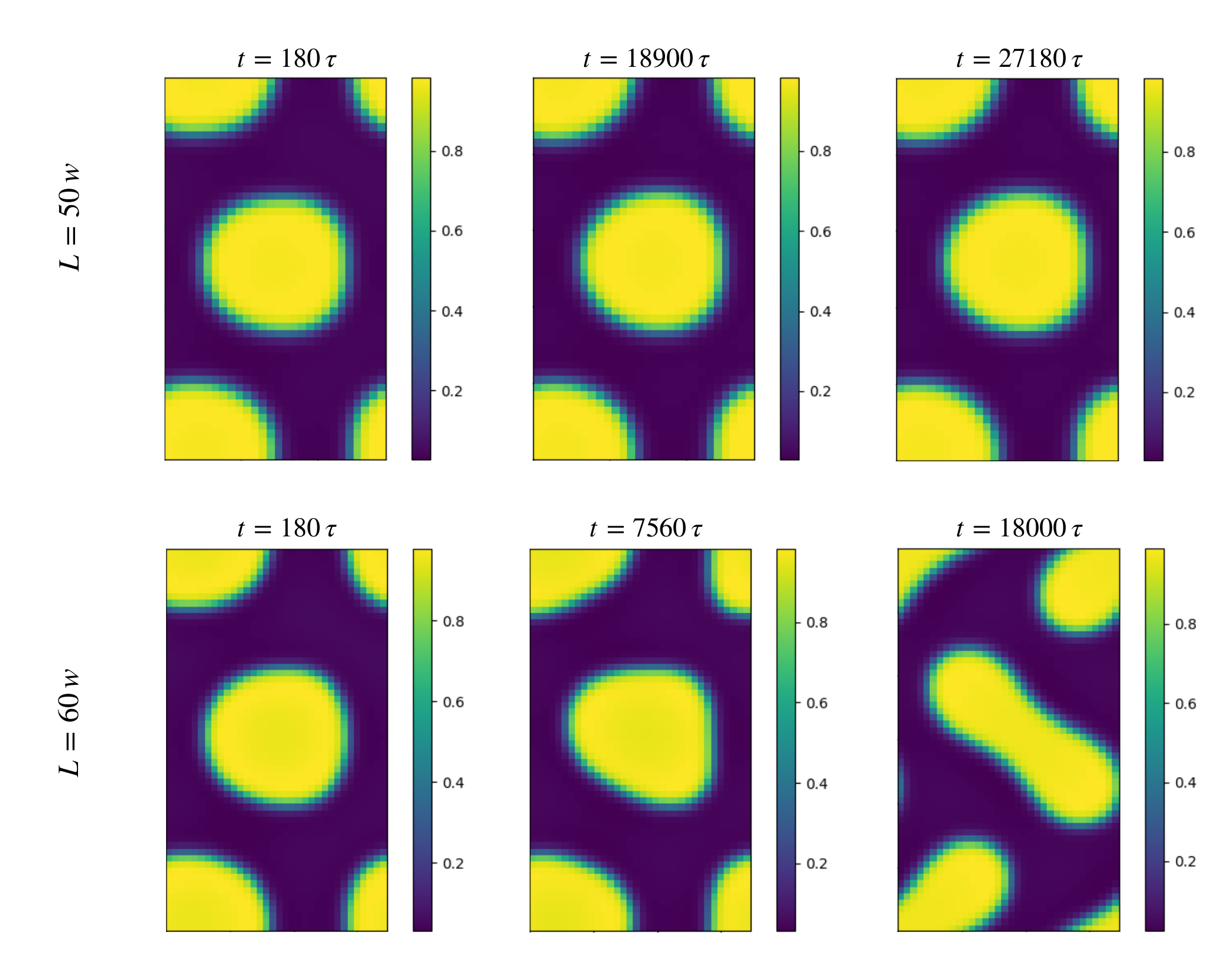}
    \caption{Snapshots of a unit cell of a hexagonal pattern for different times and two different pattern sizes. The simulation box has width $L$ and height $\sqrt{3}L$ (given in the Figure) and periodic boundary conditions to resemble the unit cell of a hexagonal pattern. Two circular droplets are initialized at positions $(L/2 \, , \ \sqrt{3}/2 L)$ and $( 0 + \delta x \, , \ 0 + \delta y) = (5\sqrt{3} \, w \, , \ 2\sqrt{3} \, w)$ to introduce an initial perturbation. Parameters: $c_0 = 0.4 \, c_\mathrm{in}$ and $k = 10^{-3}/\tau$.}
    \label{fig:ShapeDeform}
\end{figure}

The maximal size of stable droplets in a hexagonal configuration is limited by shape deformations that are induced by chemical reactions~\cite{zwickerGrowthDivisionActive2017}.
Since these shape deformations are not captured by the effective simulations, we check separately for what radii the droplets become susceptible to shape deformations.
To this end, we simulate \Eqref{eq:AdvReacCH} ($\Pe = 0$ so $\boldsymbol{v} = 0$) using a finite difference scheme (Appendix \ref{app:SimCH}) in a unit cell of the hexagonal droplet lattice and break the symmetry by shifting one of the droplets from its lattice site to induce initial disturbances.
We then distinguish between cases where the droplets exhibit growing shape instabilities (\Figref{fig:ShapeDeform} lower row) and cases where the droplet relaxes back to a circular form (\Figref{fig:ShapeDeform} upper row).

\section{Dissipation rates and predicted scaling}
\label{app:dissipation}

We measure the spatially averaged viscous dissipation rate $q_\mathrm{visc}$, which is defined as 
\begin{equation}
    q_\mathrm{visc} = \frac{Q_V}{V} = \frac{1}{V} \int \diff V \, \frac{\eta}{2} (\nabla \boldsymbol{v} + (\nabla \boldsymbol{v})^\intercal)^2
    \;,
\end{equation}
and the spatially averaged diffusive dissipation,
\begin{equation}
    q_\mathrm{diff} = \frac{Q_\mathrm{diff}}{V} = \frac{1}{V} \int \diff V \, \Lambda (\nabla \mu)^2 \; .
\end{equation}

The numerical simulations suggest that the diffusive dissipation scales as $q_\mathrm{diff}\propto t^{-2/3}$ in the late stage of passive coarsening.
To understand this scaling, we first consider a sharp interface approximation of the droplet in 2D to obtain the fluxes
\begin{align}
    j_\mathrm{in} &= 0
    & \text{and} &&
    j_\mathrm{out} &= D \frac{(c_\mathrm{out}^\mathrm{eq}(R) - \bar c)}{\ln \frac{L}{2R}} \frac{1}{r}
    \;,
\end{align}
where $c_\mathrm{out}^\mathrm{eq}(R)$ is the equilibrium concentration right outside the droplet's interface and $\bar c$ is the concentration at a distance of $L/2$ from the center of the droplet.
The integrated diffusive dissipation rate can then be approximated as
\begin{equation}
    Q_\mathrm{diff} = \sum_{i=1}^N \Lambda \int_{R_i}^{L_i/2} \diff r \, 2\pi r \, D^2 \frac{(c_\mathrm{out}^\mathrm{eq}(R_i) - \bar c)^2}{\ln^2 \frac{L_i}{2R_i}} \frac{1}{r^2} \; .
\end{equation}
Using the approximation $R_i \approx R$ and $L_i \approx L$ for all droplets $i$, we find
\begin{equation}
    Q_\mathrm{diff} = N \Lambda 2\pi D^2 \frac{(c_\mathrm{out}^\mathrm{eq}(R) - \bar c)^2}{\ln^2 \frac{L}{2R}}  \int_{R}^{L/2} \diff r \, \frac{1}{r} = N \Lambda 2\pi D^2 \frac{(c_\mathrm{out}^\mathrm{eq}(R) - \bar c)^2}{\ln^2 \frac{L}{2R}} \ln\left| \frac{L}{2R} \right| = \frac{N \Lambda 2\pi D^2 (c_\mathrm{out}^\mathrm{eq}(R) - \bar c)^2}{\ln\left| \frac{L}{2R} \right|} \;.
\end{equation}
Since $L \propto R \propto t^{1/3}$, the logarithmic term does not contribute and $(c_\mathrm{out}^\mathrm{eq}(R) - \bar{c})^2$ is constant to leading order in $t$.
However, $N \propto t^{-2/3}$ in 2D, so that we predict the whole expression to scale as $t^{-2/3}$ in two dimensions.

A similar argument holds in three dimensions, where the sharp interface approximation results in
\begin{align}
    j_\mathrm{in} &= 0
    & \text{and} &&
    j_\mathrm{out} &= D (c_\mathrm{out}^\mathrm{eq}(R) - \bar{c})\frac{R}{r^2}
    \;,
\end{align}
where again $c_\mathrm{out}^\mathrm{eq}(R)$ is the equilibrium concentration right outside the droplet's interface and $\bar c$ is the value $c$ approaches at infinity.
The integrated diffusive dissipation rate is then roughly
\begin{equation}
    Q_\mathrm{diff} = \sum_{i=1}^N \Lambda \int_{R_i}^{L_i/2} \diff r \, 4\pi r^2 \, D^2 (c_\mathrm{out}^\mathrm{eq}(R_i) - \bar{c})^2 \frac{R_i^2}{r^4} \; .
\end{equation}
Using the approximation $R_i \approx R$ and $L_i \approx L$ for all droplets $i$, we find
\begin{equation}
    Q_\mathrm{diff} \approx N \Lambda  4\pi\, D^2 (c_\mathrm{out}^\mathrm{eq}(R) - \bar{c})^2 \int_R^{L/2} \diff r \, \frac{R^2}{r^2} \; .
\end{equation}
Performing the integral, we get
\begin{equation}
    Q_\mathrm{diff} \approx N \Lambda  4\pi\, D^2 (c_\mathrm{out}^\mathrm{eq}(R) - \bar{c})^2 \left(R - \frac{2R^2}{L}\right)
    \;,
\end{equation}
where $N \propto t^{-1}$, $(c_\mathrm{out}^\mathrm{eq}(R) - \bar{c})^2$ is constant to leading order, and $(R - 2R^2/L) \propto t^{1/3}$.
Therefore, the whole expression scales as $t^{-2/3}$ also in three dimensions.

\begin{figure}
    \centering
    \includegraphics[width=0.8\textwidth]{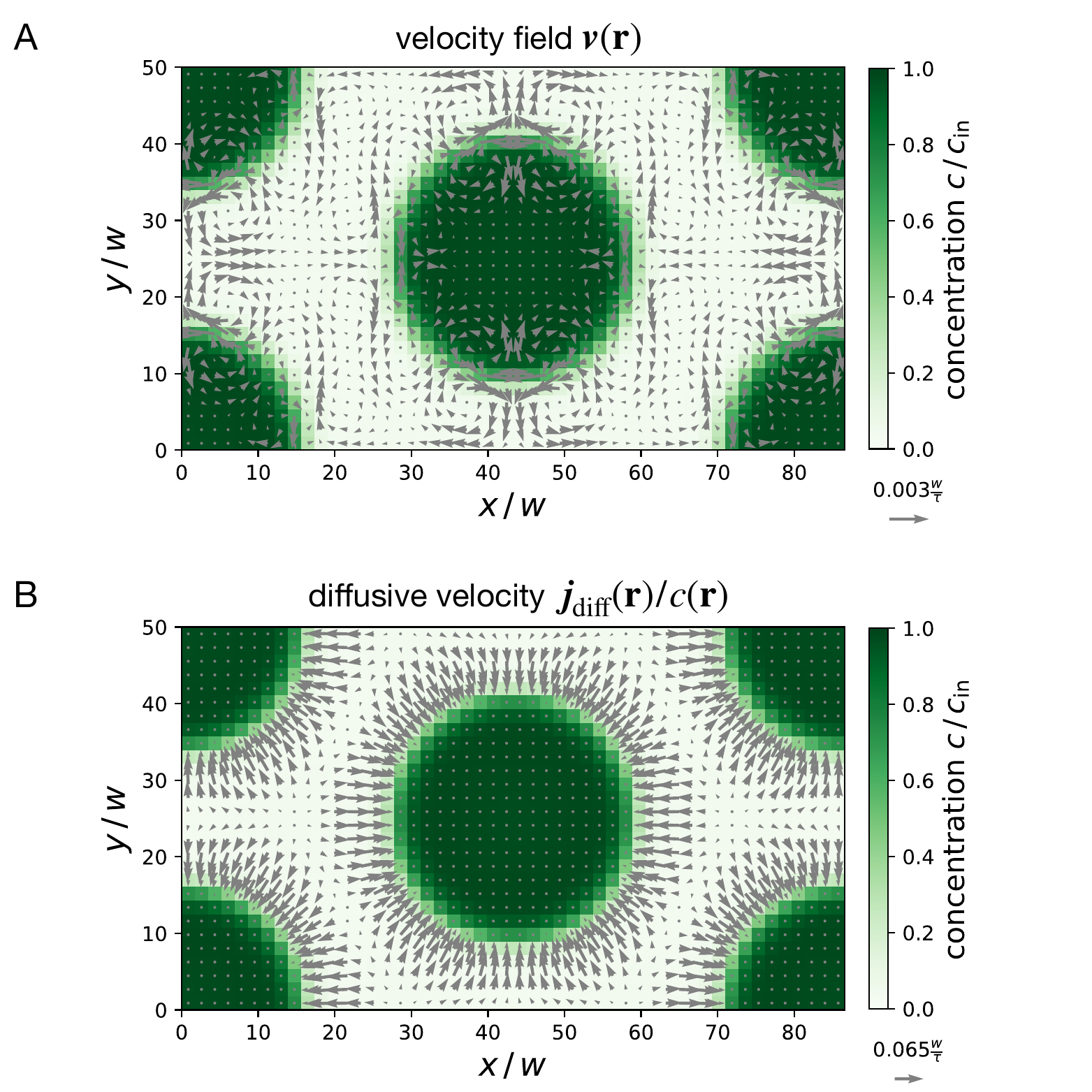}
    \caption{(A) Hydrodynamic velocity field $\boldsymbol{v}$.
    (B) Diffusive flux velocity field $\boldsymbol{j}_\mathrm{diff}/c$.
    (A--B) Both snapshots show the stationary state in a hexagonal unit cell for $k = 10^{-3}/\tau$, $c_0 = 0.4 \, c_\mathrm{in}$, and $\Pe = 100$. The legend in the lower right corner of each panel indicates that diffusive velocities are much larger, explaining why viscous dissipation is around 4 orders of magnitude smaller than the diffusive dissipation (ratio after $T = 900 \tau$ is $q_\mathrm{visc} / q_\mathrm{diff} = 2.30 \cdot 10^{-4}$).
    }
    \label{fig:dissipation_velField}
\end{figure}

\end{document}